\documentclass[fleqn,usenatbib,useAMS]{mnras}
\usepackage{graphicx}   
\usepackage{amsmath}    
\usepackage{amssymb}    
\usepackage{multicol}        
\usepackage{bm}         
\usepackage{pdflscape}
\usepackage[T1]{fontenc}
\usepackage{ae,aecompl}

\usepackage{newtxtext,newtxmath}

\def\oc3{[O~{\sc iii}]$_c$}
\def\ob3{[O~{\sc iii}]$_b$}

\title[Mrk1018]{Double tidal disruption events in the changing-look AGN Mrk1018}

\author[Zhang X. G.]{Xue-Guang Zhang$^{1}$\thanks{Contact e-mail: \href{mailto:xgzhang@gxu.edu.cn}{xgzhang@gxu.edu.cn}}\\
$^{1}$Guangxi Key Laboratory for Relativistic Astrophysics, School of Physical Science and Technology,
GuangXi University, Nanning, 530004, P. R. China}

\date{}
\begin{document}
\label{firstpage}
\pagerange{\pageref{firstpage}--\pageref{lastpage}}
\maketitle

\begin{abstract}
Tidal disruption events (TDEs) as excellent beacons to black hole (BH) accreting systems have been studied for more than five decades 
with a single star tidally disrupted by a central massive BH. However, if considering two stars passing through a central BH and being 
tidally disrupted in a short period, so-called double TDEs could be expected and lead to unique variability features very different 
from features from standard TDEs. Here, we report such oversimplified double TDEs in the known changing-look AGN Mrk1018, of which 
15years-long light curve with plateau features can be described by two main-sequence stars tidally disrupted by the central supermassive 
BH. Meanwhile, the BH mass determined by the double TDEs is consistent with the M-sigma relation determined value by the measured 
stellar velocity dispersions in Mrk1018. The results indicate tight connections between the TDEs and the changing-look properties in 
Mrk1018. 
\end{abstract}


\section{Introduction}

	A star can be tidally disrupted by gravitationally tidal force of a massive black hole (BH), when it passing close to the 
central BH with a distance larger than the central BH event horizon but smaller than the tidal disruption radius, leading to the 
so-called tidal disruption events (TDEs) as the best beacons for BH accreting systems as discussed in \citet{ref1, ref2, ref3, ref4, 
ref5, ref6, ref7, ref8, ss24}. Moreover, accreting fallback materials to central BH in a TDE can lead to observable flare-up followed 
by a smooth decline trend in multiple wavelength bands. Based on the TDEs expected time-dependent long-term variability patterns, 
there are around 300 TDEs reported in the literature \citep{ref9, ref10, ref11, ref12, ref13, ref14, ref15, ref16, ref17, ref18, yr23}.

	Although TDEs are quickly becoming commonplace, only the case is considered with a single star tidally disrupted by central 
massive BH in each TDE. However, as known in our Galaxy in \citet{ref19}, there are tens of different types of stars around the central 
BH. Meanwhile, theoretical simulations in \citet{ref20} have shown considerable star-formations in central regions around massive BHs, 
when giant molecular clouds are being accreted. In consequence, central regions not only in quiescent galaxies but also in Active Galactic 
Nuclei (AGN) can provide efficient surroundings with numbers of stars in small volume around central BHs, and then provide chances 
for more than one stars tidally disrupted.

	Once two stars were tidally disrupted by a central BH in a short period, called as double TDEs in this paper, there could be 
more interesting properties on expected long-term variability with probable two apparent peaks, especially as the results discussed 
in \citet{ref21}, far away from the standard TDEs expected variability patterns. It will be very interesting to detect such double 
TDEs. Here, we report that the long-term photometric variability of the known changing-look AGN Mrk1018 can be described by the 
oversimplified double TDEs.

\begin{figure*}
\centering\includegraphics[width = 18cm,height=14cm]{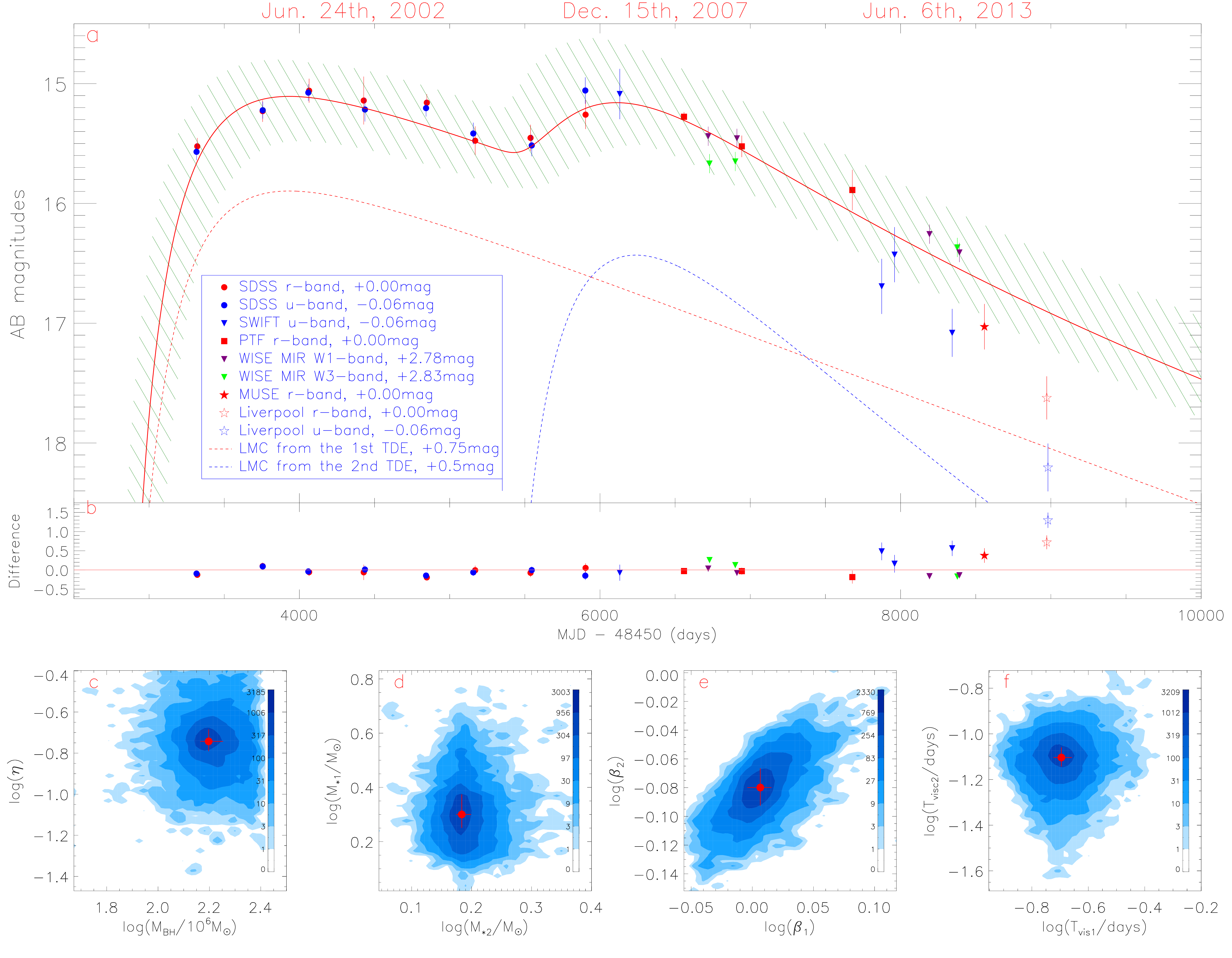}
\caption{Panel a shows the long-term light curve of Mrk1018 after subtractions of the host galaxy contributions, and the best 
descriptions by the double TDEs. As shown in the Legend, solid red circles show the data points in the r-band from the SDSS 
stripe82 database, solid blue circles show the data points minus 0.06 magnitudes in the u-band from the SDSS stripe82 database, solid 
blue triangles show the data points minus 0.06 magnitudes in the u-band from the SWIFT telescope, solid red squares show the data 
points in the r-band from the PTF, solid purple triangles show the data points plus 2.78 magnitudes in the MIR W1 band from the WISE, 
solid green triangles show the data points plus 2.83 magnitudes in the MIR W2 band from the WISE, solid red five-point-star shows the 
data point in the r-band from the MUSE telescope, open red five-point-star shows the data point in the r-band from the Liverpool 
telescope, open blue five-point-star shows the data point minus 0.06 magnitudes in the u-band from the Liverpool telescope. The solid 
red line and areas covered by green lines show the best descriptions and the corresponding confidence bands with accepted 1$\sigma$ 
uncertainties of the model parameters. Here, in order to show clean results with only one solid red line as the best fitting results, 
the data points from different wavelength bands are shifted as shown in the Legend through the best fitting results to light curves 
in different bands. Panel b shows the difference between the observed light curve and the best descriptions, with the horizontal 
solid red line shows the difference equal to zero. Panel c, d, e, f show the two-dimensional projections of the posterior distributions 
of the model parameters applied in Mrk1018. The solid circles plus error bars show the accepted values and the corresponding 1$\sigma$ 
uncertainties of the model parameters. Different number densities relative to different colours are shown in the colour bar in the 
right region of each bottom panel.}
\label{tde}
\end{figure*}

        This paper is organized as follows. Section 2 presents our main results and necessary discussions on the long-term variability 
of Mrk1018 through the oversimplified double TDEs. Section 3 gives our main conclusions. And in this paper, we have adopted the 
cosmological parameters of $H_{0}=70{\rm km\cdot s}^{-1}{\rm Mpc}^{-1}$, $\Omega_{\Lambda}=0.7$ and $\Omega_{\rm m}=0.3$.

\section{Main Results and Discussions} 

	Mrk1018 is a well-known changing-look AGN as discussed in \citet{ref22}, with its type changed from Seyfert 1.9 in 1981 to 
Seyfert 1 in 1985, to broad line quasar in SDSS in 2000, and then to a Seyfert 1.9 in 2015. The long-term light curve from 2001 to 
2015 is shown in the panel a of Fig.~\ref{tde}, after subtractions of the host galaxy contributions as reported in \citet{ref22}. 

	Due to the apparent lasting plateau features in the long-term light curves, an assumed classical TDE (a single star tidally 
disrupted) had been rejected in \citet{ref22}. However, the long-term variability features from 2011 to 2015 show a smooth decline 
trend which is not common for variability of broad line AGN but can be simply described by assumed TDEs, indicating probable 
contributions from TDEs in Mrk1018. Meanwhile, there are two apparent peaks around Jun 2002 and Jan 2008\ in the long-term light 
curve of Mrk1018. Therefore, the proposed double TDEs model can be considered in Mrk1018, with not a single star but two stars 
tidally disrupted by the central massive BH. 

	It is interesting to consider whether double TDEs can be applied to describe the long-term variability properties of Mrk1018, 
especially the previously reported peculiar lasting plateau features (or the two apparent peaks). Based on the known standard 
theoretical TDE model \citep{ref16} and the corresponding public codes of TDEFIT/MOSFIT \citep{ref7} and also what we have 
recently done on TDEs candidates in \citet{zh22a, zh22b, zh23a, zh23b, zh24, zh24b, zh25}, oversimplified double TDEs can be simply 
applied by the following three steps.

	First, as discussed in \citet{ref7}, viscous-delayed accretion rates $\dot{M}_a$ can be determined by 
\begin{equation}
	\dot{M}_a(T_{vis}, \beta, t)~=~\frac{exp(-t/T_{vis})}{T_{vis}}\int_{0}^{t}exp(t'/T_{vis})\dot{M}_{fbt}dt'
\end{equation}
through TDEFIT/MOSFIT provided fallback material rates $\dot{M}_{fbt}$ for standard cases with a $M_{*}=1{\rm M_\odot}$ main sequence 
star tidally disrupted by a central massive BH with mass $M_{BH}=10^6{\rm M_\odot}$ with different impact parameters $\beta$ and 
different viscous delay times $T_{vis}$.  

	Second, for double TDEs with input model parameters of $M_{\rm BH}$ and $M_{*,1}$ and $M_{*,2}$ (stellar masses of the two 
stars) different from $10^6{\rm M_\odot}$ and $1{\rm M_\odot}$, the actual viscous-delayed accretion rates $\dot{M}_1$ and 
$\dot{M}_2$ of two separated TDEs can be created by the following scaling relations,
\begin{equation}
\dot{M}_i(t_i) = M_{\rm BH,6}^{-0.5} M_{\star,i}^2 R_{\star,i}^{-1.5} \dot{M}_{a}(T_{vis,i}, \beta_i, t_i) \ \ \ (i=1,2) \\
\end{equation}
with $M_{\rm BH,6}$, $M_{\star,i}$, $R_{\star,i}$ as central BH mass in units of ${\rm 10^6M_\odot}$, stellar masses in units of 
${\rm M_\odot}$ and stellar radii in units of ${\rm R_{\odot}}$ of the two stars, respectively. And the mass-radius relation in 
\citet{ref24} has been accepted for main-sequence stars. For double TDEs, joint time-dependent viscous-delayed accretion rates can 
be simply described by $\dot{M}(t)=\dot{M}_1(t) + \dot{M}_2(t + \Delta t)$ with $\Delta t$ as starting date interval between the 
two separated TDEs.

	Third, based on the bolometric luminosity from the joint time-dependent viscous-delayed accretion rates and the AB magnitudes 
of the sun, the time dependent AB magnitudes in one applied filter can be simply estimated by 
\begin{equation}
\begin{split}
&L_{bol}(t) = \eta \dot{M}(t) c^2\\
&mag(t) = mag_{\odot} - 2.5\log(\frac{L_{bol}(t)}{3.83\times10^{33}{\rm erg/s}}(\frac{D_\odot}{D_M})^2)
\end{split}
\end{equation}
with $D_M$ and $D_\odot$ as the distance of Mrk1018 and the sun to the earth, $mag_{\odot}$ as the apparent AB magnitude in the 
applied filter (such as for the r-band, $mag_{\odot}=-27.08$).

\begin{figure*}
\centering\includegraphics[width = 18cm,height=5cm]{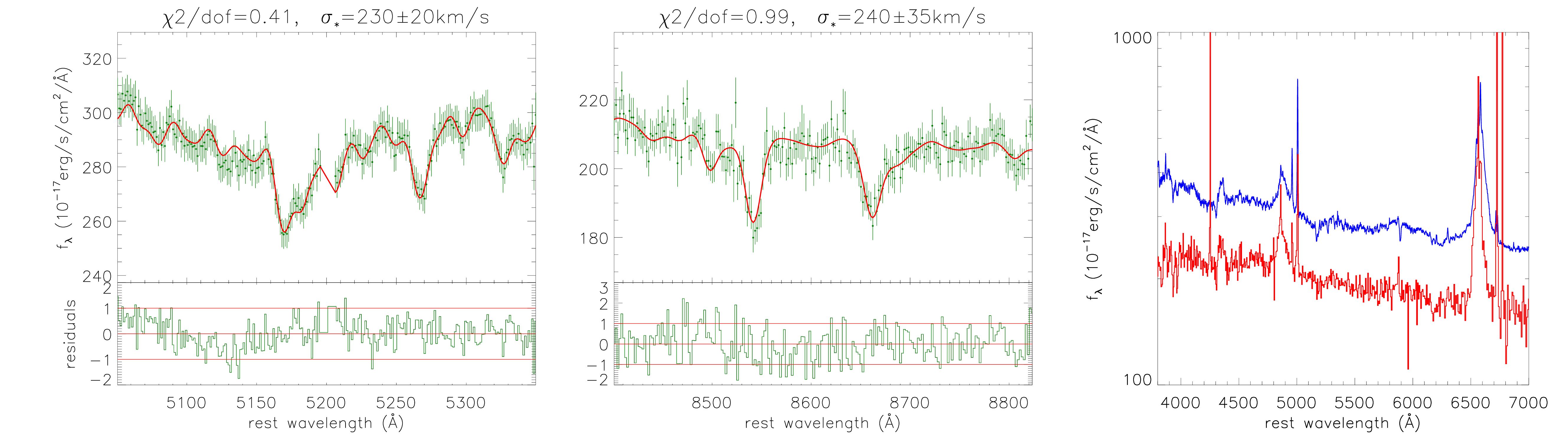}
\caption{Left and middle panel show the best fitting results (solid red line in top regions) and the corresponding residuals (bottom 
regions) to the SDSS observed spectrum (dark green circles plus error bars) around 5175\AA~ and around 8600\AA. In the bottom regions 
of left and middle panel, the residuals are calculated by the SDSS spectrum minus the best fitting results and then divided by the 
uncertainties of the SDSS spectrum, and the horizontal solid red lines show residuals$=\pm1$, respectively. Right panel 
shows the SDSS spectrum (solid line in blue) (MJD=51812) of Mrk1018 at bright phase, and the SDSS spectrum (solid line in red, scaled 
by 30) of the TDE candidate of SDSS J0159 at bright phase.}
\label{ssp}
\end{figure*}

	Based on the proposed double TDEs with free model parameters of BH mass $M_{BH}$, stellar masses $M_{*,1}$ and $M_{*,2}$ of 
two stars, ($\beta_1$, $T_{vis,1}$) and ($\beta_2$, $T_{vis,2}$) for the two separated TDEs, energy transfer efficiency 
$\eta$ and the starting date interval $\Delta t$ between the two separated TDEs, the best descriptions and the corresponding confidence 
bands to the long-term light curve of Mrk1018 can be determined and shown in the panel a of Fig.~\ref{tde}, through the maximum 
likelihood method combining with the Markov Chain Monte Carlo (MCMC) technique \citep{ref23} applied with the prior uniform 
distributions of $\Delta t$ larger than 0 and smaller than 4500days (at least two times larger than the time interval for the two 
peaks in the light curves) and of the other model parameters similar as those described in \citet{ref7}. Here, in order to show clean 
results in Fig.~\ref{tde} with only one solid red line to represent the best fitting results for the r-band data points, the data 
points in the other bands are simply shifted based on the best fitting results to the data points in the other bands. Moreover, as 
discussed in \citet{br19}, collisions between two debris streams could be expected in double TDEs. Emissions from such efficient 
collisions can make flares leading the observed light curves to be un-smooth. However, as shown in Fig.~\ref{tde}, the observed light 
curves are smooth enough without steep flares. In other words, even there were collisions between two debris streams in assumed 
double TDEs in Mrk1018, the corresponding effects are tiny. Therefore, in our oversimplified double TDEs, effects of such collisions 
are not considered.

	The MCMC technique determined posterior distributions of the main model parameters for the double TDEs are shown in the bottom 
panels of Fig.~\ref{tde}, leading to the BH mass of Mrk1018 about $157_{-12}^{+23}\times10^6{\rm M_\odot}$ and the stellar mass 
$1.53_{-0.02}^{+0.06}{\rm M_\odot}$ and the corresponding stellar radius about $1.86{\rm R_{\odot}}$ determined through the mass-radius 
relation in \citet{ref24} for the first disrupted star with polytropic index $\gamma=4/3$, and the stellar mass 
$1.99_{-0.22}^{+0.37}{\rm M_\odot}$ and the corresponding mass-radius relation determined stellar radius about $2.16{\rm R_{\odot}}$ 
for the second disrupted star with polytropic index $\gamma=4/3$. The starting date interval of the two stars tidally disrupted is 
about 2350$\pm$180 days. Now, the long-term optical light curve with the two apparent peaks (or with the lasting plateau features) 
can be described by the assumed double TDEs in Mrk1018.

	Before proceeding further, as the Eqs.6 in \citet{yr23}, the determined BH mass $157\times10^6{\rm M_\odot}$ in Mrk1018 is 
smaller than the Hills limits about $226\times10^6{\rm M_\odot}$ and $247\times10^6{\rm M_\odot}$ after considering stellar parameters 
of the two stars. Meanwhile, comparing with the values of TDE model parameters for the optical TDEs in \citet{ref7}, the 
determined model parameters by the proposed double TDEs model in Mrk1018 are common and reasonable for the stellar masses, the impact 
parameters about 1.015 and 0.832 and the viscous delay time about 0.2 days and 0.08 days and the energy transfer efficiency 18.1\%.

	Furthermore, the proposed double TDEs model determined mean bolometric luminosity is about $5.9\times10^43$erg/s at bright 
phase with MJD-48450 from 4000 to 6500 in Mrk1018, simply consistent with the estimated bolometric luminosity about 
$3.1\times10^{44}$erg/s at bright phase in Mrk1018 in \citet{ref22} through SDSS spectroscopic features. Meanwhile, the proposed double 
TDEs model can lead to about 0.83${\rm M_\odot}$ (54\% of the total stellar mass) and 0.30${\rm M_\odot}$ (15\% of the total stellar 
mass) stellar masses accreted by the first and second disruption event. The corresponding ratios of accreted stellar mass to the 
total stellar mass are common as the values for the optical TDEs listed in Table~6 in \citet{ref7}.

\begin{figure}
\centering\includegraphics[width = 9cm,height=6cm]{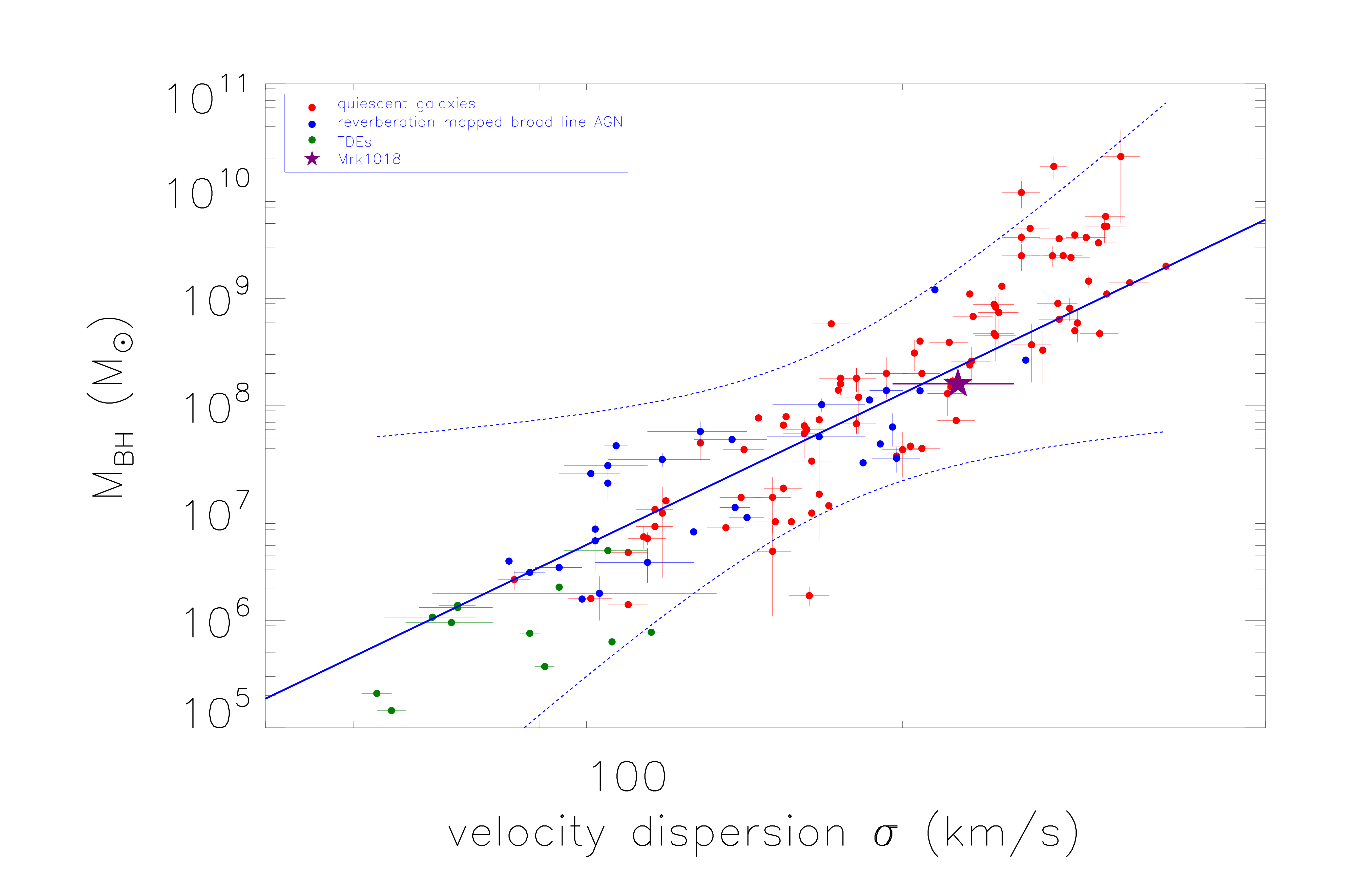}
\caption{On the properties of Mrk1018 in the space of BH mass versus stellar velocity dispersion. As shown in the Legend, solid blue 
circles plus error bars show the results for the 89 quiescent galaxies in \citet{ref37} with the BH masses determined by dynamic 
methods, solid red circles plus error bars shows the results for the 29 reverberation mapped broad line AGN in \citet{ref38} with 
the determined viral BH masses, solid dark green circles plus error bar shows the results for the 12 TDEs in \citet{ref39}, solid 
purple five-point-star show the results of Mrk1018 with the BH mass determined by the double TDEs. Solid blue line and dashed blue 
line shows the more recent M-sigma relation in \citet{ref40} and the corresponding 5$\sigma$ confidence bands determined by the 
F-test technique.}
\label{msig}
\end{figure}

	Besides the determined central BH mass by the double TDEs model applied in  Mrk1018, the central BH mass of Mrk1018 can also 
be determined by the M-sigma relation \citep{ref25, ref26} through measured stellar velocity dispersion of the host galaxy. Left 
and middle panel of Fig.~\ref{ssp} show the best descriptions determined by the commonly applied Simple Stellar Population (SSP) 
method \citep{ref27, ref28} to the absorption features of Mg~{\sc i} and Ca~{\sc ii} triplet in the quasar-like spectrum 
(plate-mjd-fiber=404-51812-141) of Mrk1018 in Sloan Digital Sky Survey (SDSS), considering the 39 stellar templates \citep{ref28} 
plus a power law continuum emission component, leading the measured stellar velocity dispersion to be about 235$\pm$35km/s in 
Mrk1018. Considering the measured stellar velocity dispersion, the double TDEs model determined BH mass is consistent with the 
M-sigma relation estimated BH mass in Mrk1018, as shown in Fig.~\ref{msig}, providing further clues to support the double TDEs 
in Mrk1018. Meanwhile, Comparing with the reported spectroscopic features of optical TDEs, we can find that the SDSS 
spectrum of Mrk1018 at bright phase is totally similar as the SDSS spectrum (plate-mjd-fiber=403-51871-549) of the TDE candidate 
of SDSS J0159 (also one known changing-look quasar discussed in \citealt{lc15}) at the brigh phase as discussed in \citet{ref15}. 
Therefore, there are no clues to support unique spectroscopic features in the Mrk1018 at the bright phase.

\begin{figure}
\centering\includegraphics[width = 8.5cm,height=4.5cm]{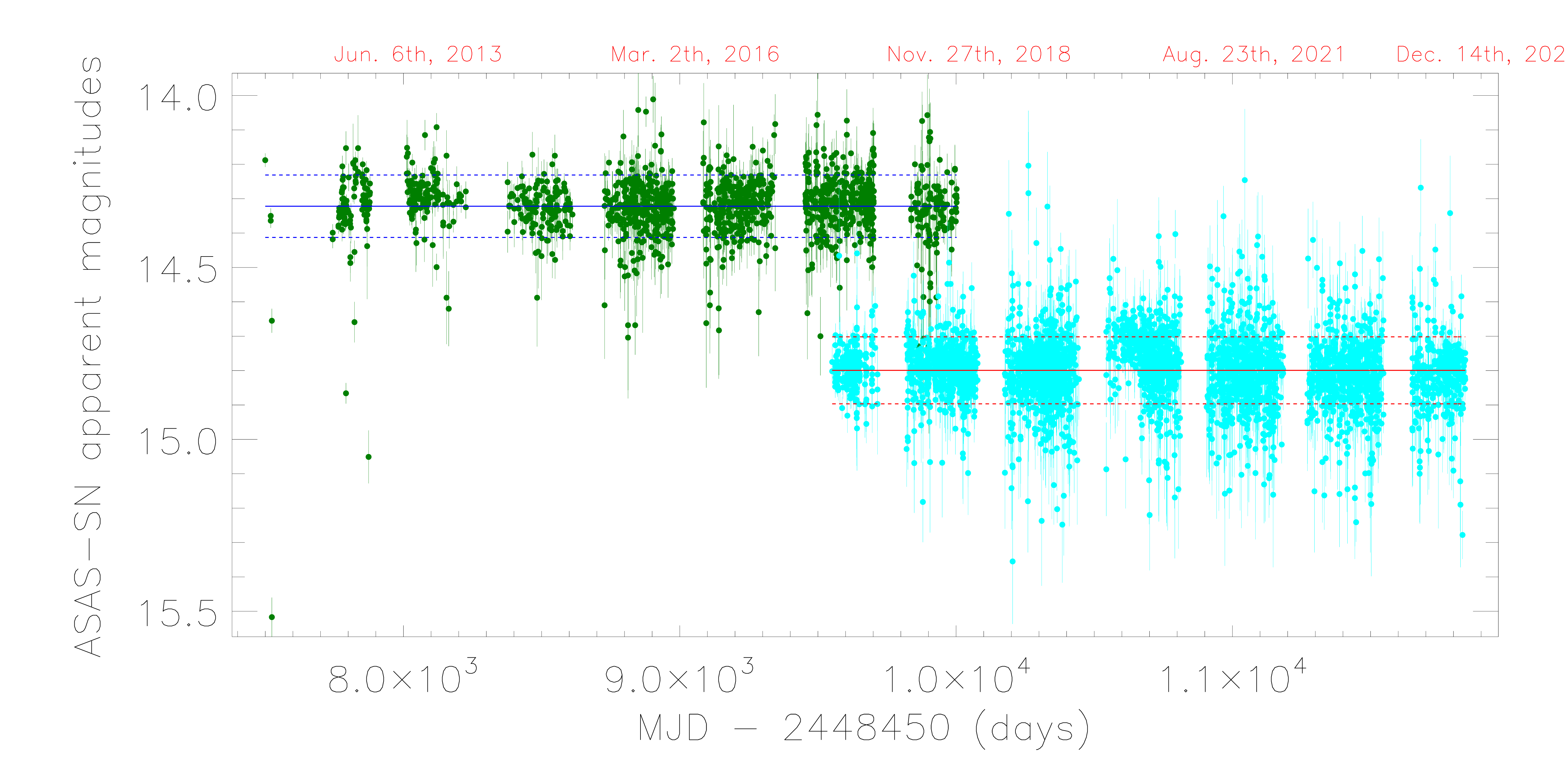}
\caption{The V-band and g-band light curves of Mrk1018 from the ASAS-SN. Solid dark green circles plus error bars show the data 
points in the V-band. Horizontal solid blue line and dashed blue lines mark the mean magnitude of the V-band light curve and the 
corresponding scatters of the standard deviations. Solid cyan circles plus error bars show the data points in the g-band. Horizontal 
solid red line and dashed red lines mark the mean magnitude of the g-band light curve and the corresponding scatters of the standard 
deviations.
}
\label{lmc}
\end{figure}

	Moreover, the following up-to-today long-term V-band and g-band light curves including the host galaxy contributions of 
Mrk1018 are collected from the All-Sky Automated Survey for Supernovae (ASAS-SN) \citep{ref29, ref30} from Jan 23 2012 to Dec 14 
2023, and shown in Fig.~\ref{lmc}. For the V-band light curve from Jan 23 2012 to Jan 6 2018, the mean apparent magnitude is about 
14.33mag with the corresponding standard deviation about 0.091mag. For the g-band light curve from Sep 5 2017 to Dec 14 2023, the 
mean apparent magnitude is about 14.80mag with the corresponding standard deviation about 0.097mag. The 12years-long very weak 
variability in the ASAS-SN light curves can provide clues to support that the apparent variability shown in the panel a of 
Fig.~\ref{tde} is not due to the commonly accepted intrinsic AGN variability in Mrk1018. 

	Meanwhile, accepted the double TDEs applied in Mrk1018, the determined $\beta1\sim1.02$ and $\beta_2\sim0.83$ are smaller 
than the critical value around 2.0 as discussed in \citet{ref6, sp24}, indicating the corresponding two TDEs to be partial TDEs. 
As partial TDEs discussed in \citet{so25}, the remainder after the first disruption can be re-disrupted for a second bright 
phase. However, as the shown light curves from ASAS-SN, there are no re-brightening phase in Mrk1018. Meanwhile, as determined 
stellar parameters disrupted by the proposed double TDEs, the second TDE should have its star with stellar mass about 
1.99${\rm M_\odot}$ totally larger than the mass only about $1.53-0.83=0.7{\rm M_\odot}$ of the reminder of the first TDE, therefore, 
the re-brightening partial TDEs scenario is not preferred for the flare between 2001 and 2015 in Mrk1018. That is the main reason for 
application of the proposed double TDEs, rather than the re-brightening partial TDEs.

	For Mrk1018 as a changing-look AGN, there is no unambiguous physical mechanism to explain the changing-look spectroscopic 
properties. However, as discussed in \citet{ref22}, Mrk1018 has its broad H$\alpha$ with the line width (full width at half maximum) 
of 4000km/s in the Type-1 like SDSS spectrum changed to about 3300km/s in the Type-1.9 like Multi-Unit Spectroscopic Explorer (MUSE) 
spectrum, against the Virialization assumption \citep{ref31} expected results that the broad line width increases as the AGN luminosity 
falls. However, once accepted TDEs in Mrk1018, the unique variability properties of the broad H$\alpha$ can be explained by the 
expanding emission line structures over time but the weakening line emissions, similar as the variability properties of the broad 
H$\alpha$ in the TDE ASASSN-14li as shown in \citet{ref32}. Therefore, as discussed in \citet{ref22}, the double TDEs determined BH 
mass (or the M-sigma relation determined BH mass) are about 2 times larger than the virial BH mass at the high state and about 12 
times larger than the virial BH mass at the low state, due to the central BLRs being not virialized in Mrk1018. As a changing-look 
AGN, the double TDEs in Mrk1018 can be applied to explain both the long-term photometric variability properties and the multi-epoch 
optical spectroscopic variability properties, providing strong evidence to support that the central double TDEs are tightly connected 
to the changing-look properties in the changing-look AGN Mrk1018.

	Before ending the section, as more recent discussions in \citet{yr23} that the average optical TDE rate is about $10^{-5}$ 
per galaxy per year, this indicates the rate for such double TDEs should be extremely lower than $10^{-10}$ per galaxy per year even 
without considerations of geometric structures of two stars in space around central BHs in galaxies. However, it is precisely because 
the detection rate for such double TDEs is extremely low, it is meaningful to detect and discuss such double TDEs which can lead to 
observed light curves having double-peaked features. Therefore, Mrk1018 is collected as the subject of the paper. Moreover, the best 
descriptions to the observed light curve of Mrk1018 by double TDEs can not be accepted as the stable evidence to confirm such double 
TDEs in Mrk1018. However, besides the detection rate for double TDEs and the model determined best descriptions, the consistent BH 
mass by the model with the mass by the M-sigma relation and the explanations to the line width variability of broad emission line 
after considerations of TDEs contributions can be applied to support such double TDEs in Mrk1018. In other words, the double TDEs 
can be accepted as a preferred model in the changing-look AGN Mrk1018, but further efforts are necessary to support such double TDEs 
in Mrk 1018.

	Furthermore, considering the double TDEs with two stars tidally disrupted around the central supermassive BH in the 
changing-look AGN Mrk1018, it is more interesting to expect similar double TDEs in some kpc-scale dual galaxies (or kpc-scale dual 
AGNs), such as the dual systems expected by the optical double-peaked features in the narrow emission lines discussed in \citet{ref33, 
ref34, ref35, ref36} but without spatially resolved two cores in photometric images. One star is tidally disrupted in one galaxy of 
a dual system, but the other star is tidally disrupted in the other galaxy in the dual system, and the two stars are disrupted 
without any influence on each other, leading to more flexible variability patterns (far deviating from the standard TDEs expected 
patterns) for the double TDEs in such kpc-scale dual galaxy systems. 

\section{Conclusions}

	The scenario on two stars tidally disrupted by central supermassive BH as double TDEs is proposed to explain the long-term 
photometric variability properties of Mrk1018 with two bright peaks followed by a smooth decline trend. Moreover, the double TDEs 
determined BH mass of Mrk1018 is well consistent with M-sigma relation expected value, but very different from the virial BH mass 
of Mrk1018 determined by properties of broad Balmer emission lines, providing clues to support the double TDEs in Mrk1018.

\section*{Acknowledgements}
Zhang gratefully acknowledge the anonymous referee for giving us constructive comments and suggestions to greatly 
improve the paper. Zhang gratefully thanks the kind financial support from GuangXi University and the kind grant support from 
NSFC-12173020 and NSFC-12373014, and the support from Guangxi Talent Programme (Highland of Innovation Talents). This paper has 
made use of the NASA/IPAC Extragalactic Database (NED). This paper has made use of the data from the ASAS-SN, and from the SDSS. 
This paper has made use of the public code of TDEFIT, MOSFIT, and emcee package.

\section*{Data Availability}
The data underlying this article will be shared on reasonable request to the corresponding author
(\href{mailto:xgzhang@gxu.edu.cn}{xgzhang@gxu.edu.cn}).

\label{lastpage}
\end{document}